\documentclass[preprint,12pt]{elsarticle}

\usepackage{bbding}
\usepackage{xcolor}
\usepackage{multirow}

\usepackage{amssymb}
\usepackage{hyperref}
\usepackage{float}

\journal{SoftwareX}

\begin{document}

\begin{frontmatter}



\title{EASELAN: An Open-Source Framework for Multimodal Biosignal Annotation and Data Management}


\author{Rathi Adarshi Rammohan, Moritz Meier, Dennis Küster, Tanja Schultz}
\affiliation{organization={Cognitive Systems Lab, University of Bremen}, \newline 
            addressline={Enrique-Schmidt-Str.~5}, 
            city={Bremen},
            postcode={28359}, 
            state={Bremen},
            country={Germany}}



\begin{abstract}
Recent advancements in machine learning and adaptive cognitive systems are driving a growing demand for large and richly annotated multimodal data. A prominent example of this trend are fusion models, which increasingly incorporate multiple biosignals in addition to traditional audiovisual channels. This paper introduces the EASELAN annotation framework to improve annotation workflows designed to address the resulting rising complexity of multimodal and biosignals datasets. It builds on the robust ELAN tool by adding new components tailored to support all stages of the annotation pipeline: From streamlining the preparation of annotation files to setting up additional channels, integrated version control with GitHub, and simplified post-processing. EASELAN  delivers a seamless workflow designed to integrate biosignals and facilitate rich annotations to be readily exported for further analyses and machine learning-supported model training. The EASELAN framework is successfully applied to a high-dimensional biosignals collection initiative on human everyday activities (here, table setting) for cognitive robots within the DFG-funded Collaborative Research Center 1320 Everyday Activity Science and Engineering (EASE). In this paper we discuss the opportunities, limitations, and lessons learned when using EASELAN for this initiative. To foster research on biosignal collection, annotation, and processing, the code of EASELAN is publicly available\footnote{ https://github.com/cognitive-systems-lab/easelan}, along with the EASELAN-supported fully annotated Table Setting Database\cite{HartmannMeierPutzeSchultz2025}.
\end{abstract}

\begin{highlights}
\item EASELAN is a comprehensive human-in-the-loop framework for high-dimensional sensor data annotation aligning multiple modalities including audio, text, video, and biosignals resulting from human behavior and their activities of the eyes, the body, the skin, the muscles, and the brain.
\item EASELAN builds upon the well-established ELAN-tool to streamline integration, synchronization, formatting, processing, transcription, annotation, access, sharing, and preservation of signals from audio, text, video, and biosignals.
\item EASELAN is an open-source framework that leverages Git for version control and Continuous Integration/Continuous Deployment
pipelines for annotation, validation and automated feedback.
\end{highlights}


\end{frontmatter}



\section{Motivation and Significance}
\label{sec:general}
With the increasing demand of machine learning methods and biosignal-adaptive systems for detailed and holistic perspectives on human behavior \cite{schultz_biosignals_2023}, the need for powerful tools to annotate high-dimensional biosignals becomes more pressing.
State-of-the-art annotation tools, such as ELAN \cite{wittenburg-etal-2006-elan} and ANVIL \cite{kipp_anvil_2001} offer a variety of features for multimodal annotations. However, most of these tools were originally designed to annotate and visualize audiovisual data rather than heterogeneous data formats and high-dimensional multimodal biosignals. Furthermore, most tools are not prepared to support a semi-automated annotation process with a human-in-the-loop design supported by dedicated machine learning-based annotation models. 

Today, advanced smart devices already capture large and heterogeneous data sets \cite{naghib_comprehensive_2023}. Likewise, researchers face new challenges in managing data collected across multiple or interconnected labs \cite{schultz_lablinking_2024}, where distributed annotation is becoming the norm rather than the exception.
In addition, modern biosignal-adaptive systems leverage highly complex biosignals to dynamically adapt to user needs in real-time applications \cite{schultz_biosignals_2023, kothe2025lab} and thus offer powerful machine learning-based models that might be useful for semi-supervised data annotation. Together, these developments call for several advances in multimodal annotation tools. The open-source framework EASELAN \cite{MeierRammohanSchultz2025} aims to fill this gap with optimized workflows for integration, synchronization, formatting, processing, transcription, annotation, access, preservation, and dissemination of complex multimodal data from audio, video, text, and biosignals.

\subsection{Need for annotated Biosignals} 
The expanding capacity to record multimodal biosignals data raises the question of how to leverage this complex information and make it more accessible to a wide range of fields. Biosignals resulting from human behavior and activities of the body, muscles, eyes, heart, and brain are typically processed using machine learning (ML) methods \cite{kothe2025lab}, including deep learning \cite{zafar_reviewing_2023} and Large Language Models (LLMs) \cite{liu2023largelanguagemodelsfewshot}.
Examples of such biosignals are captured by inertial sensors (body motion), electromyography (muscle activity), cameras and infrared sensors (eyegaze activity), electrocardiography (heart activity) and electroencephalography (brain activity), to name just a few. 
Recent works have used biosignal data, for example, to develop and evaluate recommendation systems for people with dementia \cite{steinert_evaluation_2022,steinert_predicting_2022}, in education, where they enable empathy-driven robotic tutors \cite{obaid_endowing_2018}, and in cognitive robotics, where they improve human-robot interaction \cite{meier_comparative_2019}. Furthermore, since individual biosignals may only tell part of the picture, information from complementary biosignals is fused to improve model performance \cite{muhammad_comprehensive_2021}. Other approaches focus on improving performance by synthesizing data \cite{dablain_deepsmote_2023} or employing algorithms such as Synthetic Minority Oversampling Techniques to improve model performance for small or imbalanced data \cite{fernandez_smote_2018}, particularly in health and care applications. 

To achieve reliable and robust model performance, the acquisition of high-quality, well-labeled biosignals data remains an essential and crucial prerequisite \cite{lee_artificial_2024}. Unfortunately, it can be very challenging, time- and cost-consuming to obtain and validate accurately labeled data. In addition to substantial annotation work, labor costs, and time-to-readiness, major issues also include data heterogeneity and privacy concerns when fusing biosignals with video \cite{lee_artificial_2024}, audio, and textual data. Faced with these concerns, researchers often focus on a limited set of isolated biosignals rather than developing a comprehensive, manually annotated database. As a result, modality fusion approaches tend to prioritize easily recorded biosignals over rich multimodal data with audiovisual-context-based annotations \cite{derdiyok_biosignal_2023}. 

\subsection{Demand for Advanced Annotation Tools}
We believe that open-source frameworks for multimodal biosignal annotation, much like existing libraries for time-series processing and feature extraction \cite{Barandasetal}, will play a key role in unlocking the potential of biosignals for gaining a holistic understanding of human behavior. Well-annotated and structured multimodal data could help account, for instance, for the audiovisual context of emotions \cite{bian_understanding_2024}, attention \cite{salous_smarthelm_2022}, stress \cite{bodaghi_multimodal_2024}, consumer behavior \cite{quiles_perez_data_2024}, and engagement of people with dementia \cite{steinert_audio-visual_2021}, to give just a few examples. 
In our view, despite their considerable success, current ML modeling and data fusion approaches still face substantial challenges with respect to data validity and interpretability, particularly for multimodal datasets collected in dynamic real-world environments. Improved multimodal annotation tools may play a crucial role in shaping how modeling of biosignals and complex multimodal data are understood within the broader picture of their audiovisual and social contexts. We expect these features to become increasingly important, especially considering the growing role of interdisciplinary collaboration in creating and disseminating multimodal datasets (e.g., \cite{olugbade_human_2022}). 
Thus, the purpose of this work is to provide an advanced annotation workflow to help unlock the full potential of biosignals. 

\subsection{Review of Available Annotation Tools}
While the challenges of collecting biosignals in the wild can serve as a highlight, efficient data management and annotation are critical for research \cite{wittenburg-etal-2006-elan}. Although many tools exist, few are specifically designed to visualize and annotate multimodal data. The following review briefly highlights the most widely used tools for large-scale multimodal data annotation and compares these tools to the proposed EASELAN framework in Table \ref{tab:compareTools}.

\begin{table}[h]
    \centering
    \footnotesize
    \begin{tabular}{|l|ccccccc|} \hline 
     \multirow{2}*{Feature / Tools} & \multirow{2}*{VIA} & \multirow{2}*{WebA} & \multirow{2}*{CVAT} & \multirow{2}*{BORIS} & \multirow{2}*{ANVIL} & \multirow{2}*{ELAN} & EASE \\ 
     & & & & & & & -LAN  \\ \hline
     \multicolumn{8}{|c|}{Data Acquisition and Collection} \\ \hline
     audio, image, video & \checkmark & - & \checkmark & \checkmark  & \checkmark & \checkmark & \checkmark \\ 
     biosignals & - & - & - & - & \checkmark & - & \checkmark \\ \hline 
     \multicolumn{8}{|c|}{Data Organization and Storage} \\ \hline
      cloud-based storage & - & - & \checkmark & - & - & - & \checkmark \\ 
     formats and structure & \checkmark & \checkmark & \checkmark & \checkmark  & \checkmark & \checkmark & \checkmark \\\hline
     \multicolumn{8}{|c|}{Data Annotation and Enrichment} \\ \hline
     annotation & \checkmark & \checkmark & \checkmark & \checkmark & \checkmark & \checkmark & \checkmark \\
     automatic annotation & \checkmark & \checkmark & \checkmark & - & - & - & \checkmark \\     
     meta-data & \checkmark & \checkmark & \checkmark & \checkmark & \checkmark & \checkmark & \checkmark \\
     synchronization & \checkmark & \checkmark & \checkmark & \checkmark & \checkmark & \checkmark & \checkmark \\
     pre-processing & - & - & - & - & - & - & \checkmark\\ 
     visualization  & \checkmark & \checkmark & \checkmark & \checkmark & \checkmark & \checkmark & \checkmark \\
     \hline
     \multicolumn{8}{|c|}{Data Access and Sharing} \\ \hline
     collaboration & \checkmark & \checkmark & \checkmark & \checkmark & - & - & \checkmark \\
     progress tracking & - & \checkmark & \checkmark & - & - & - & \checkmark \\
     versioning & \checkmark & - & \checkmark & - & - & - & \checkmark \\
     web-based & \checkmark & \checkmark & \checkmark & - & - & - & \checkmark \\
     \hline
     \multicolumn{8}{|c|}{Data Security and Privacy} \\ \hline
     access control & - & \checkmark & \checkmark & - & - & - & \checkmark \\ 
     anonymization & - & - & - & - & - & - & \checkmark \\ \hline
     \multicolumn{8}{|c|}{Data Preservation and Dissemination} \\ \hline
     documentation  & \checkmark & \checkmark & \checkmark & \checkmark & \checkmark & \checkmark & \checkmark \\
     open-source & \checkmark & \checkmark & - & \checkmark & -& - & \checkmark \\ \hline
    \end{tabular}
    \caption{Characteristics of state-of-the-art tools and EASELAN framework}
    \label{tab:compareTools}
\end{table}

\begin{description}
\item \textbf{VGG Image Annotator} \cite{via2019} is a standalone annotation tool hosted on GitLab and supports image data annotation in a two-stage process: first, computer vision algorithms generate preliminary annotations, then data are manually filtered and updated. Along with images and audio, video data can also be annotated using this tool. VIA runs in a web browser without requiring installation or setup. 
\item \textbf{WebAnno} \cite{webanno2013} 
is a web-based tool for text data annotation that supports simultaneous collaborative annotation and progress tracking. Further strengths include a monitor interface to compare annotations and inbuilt support for computing inter-rater agreement. Additionally, WebAnno offers a module for crowd-sourcing via CrowdFlower's API.
\item \textbf{CVAT - Computer Vision Annotation Tool} \cite{cvat_tool} is a web-based open-source platform for multimodal data annotation. It allows simultaneous collaborative annotation through assignment of project tasks and a version history. The task status is then tracked through predefined labels. The paid version of CVAT provides additional features such as cloud-based storage and pre-trained models for automatic annotation.
\item \textbf{BORIS} \cite{friard_boris_2016} is a highly cited free and open-source annotation tool for video data that allows project-based ethograms, which can be shared with collaborators and exported in many common formats. BORIS facilitates efficient key-based annotations of behaviors and events and includes in-built visualization of sound spectrograms.
\item \textbf{ANVIL} \cite{anvil2012}, a widely used tool for video annotation includes support for biosignals - in contrast to tools discussed so far. ANVIL enables synchronized visualization and annotation of multimodal data such as audio, video, 3D motion capture, and biosignals. Transcriptions of video events are displayed in parallel tracks and stored in independent layouts. The latest version of ANVIL has expanded the capabilities of tracks to target human behavior analysis.  
\item \textbf{ELAN - EUDICO Linguistic Annotator} \cite{elan2004} is a multimedia annotation tool that allows to create, edit, visualize, analyze, and search annotations time-aligned with video and audio data. It was developed at the Max Planck Institute for Psycholinguistics in Nijmegen, NL. While the aim is to provide a technological basis for the annotation, exploitation and documentation of multi-media recordings, ELAN is specifically designed for the analysis of languages, sign languages, and gestures. 
In ELAN annotations appear in layers called `tiers', also different working modes are customized for specific tasks. For example, annotation mode is the default for working with annotations, while transcription mode is designed to enter and edit transcriptions. ELAN furthermore provides a controlled vocabulary feature to consistently store potential annotations (e.g., tiers, types, and controlled vocabulary) as templates for a given task. ELAN uses an XML-based file extension called EUDICO Annotation Format (EAF) for annotation files \cite{elan2004}.  
\end{description}
After carefully reviewing the most relevant available annotation tools (see Table \ref{tab:compareTools}), we conclude that to date, no tool or framework exists that supports all the required features for the reliable, cost-effective, and time-saving process of multimodal data annotation and curation that must be met to unlock the full potential of biosignals for human behavior analysis and biosignal-adaptive systems.

\section{Description of the EASELAN Framework}
\label{software}
EASELAN builds on the strengths of the above-described tools. In particular, EASELAN provides new software components to enhance the familiar ELAN environment with more seamless workflows and effective pipelines that simplify, manage, and enhance annotations of rich multimodal data and heterogeneous biosignals. Towards this aim, EASELAN also integrates version control tools to help pre-process, annotate, and analyze various data modalities in part. It extends the capabilities of ELAN to display and annotate not only audio and video, but also high-dimensional biosignals. Version control via GitHub enhances interdisciplinary work and tracking of changes throughout the annotation process to help ensure data integrity and promote collaboration. The characteristics of the resulting EASELAN framework are summarized in Table \ref{tab:compareTools}.

\begin{figure}[!h]
\centering
\includegraphics[width=1.0\linewidth]{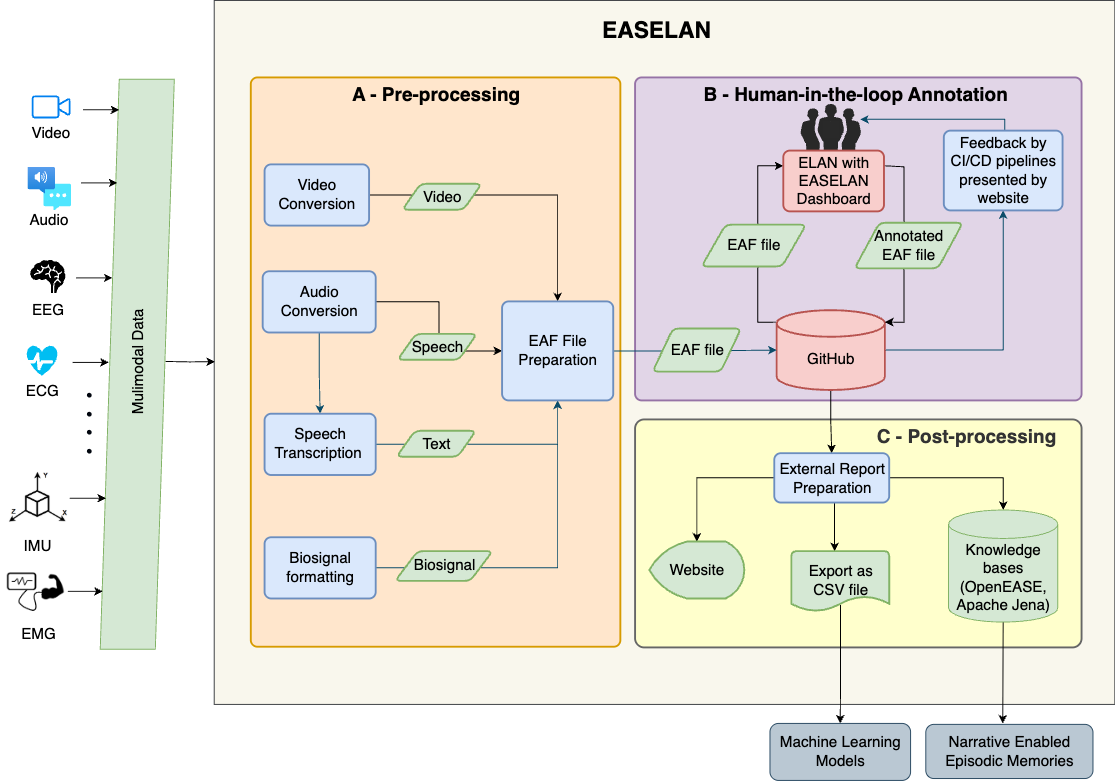}
\caption{\label{fig:pipeline}The EASELAN Architecture: (A) Pre-processing component, where the multimodal data input including high-dimensional biosignals are preprocessed and stored in EAF files. (B) Human-in-the-loop Annotation component, where EAF files are synchronized with GitHub for version control and accessed by users to annotate the data with ELAN. These annotations are validated through continuous integration and deployment (CI/CD) pipelines. (C) Post-processing component, where the EAF files together with the validated results are exported into multiple formats that can be further extended for example, to training and validating machine learning models or to retaining Narrative Enabled Episodic Memories (NEEMS), see section 3.}
    \end{figure}
    
The EASELAN architecture has three major components as depicted in Figure \ref{fig:pipeline}. 
The input to the architecture consists of multimodal data including audio, video, and biosignals resulting from measurements, for example, of brain activity by Electroencephalography (EEG), of heart activity by Electrocardiography (ECG), of body activity by inertial motion units (IMU), and of muscle activity by Electromyography (EMG), to name a few. The raw data are fed into the EASELAN Pre-processing componenent (Box A in Fig \ref{fig:pipeline}). Here, the raw data are pre-processed with dedicated conversion tools into ELAN-compatible formats. Also, speech recognition is applied to transcribe speech data, and biosignals are formatted into CSV files, see details in section \ref{pre-processing}.

The resulting data streams are stored in an EAF file to be further processed by the EASELAN Human-in-the-loop Annotation component (Box B in Fig \ref{fig:pipeline}). Here, the EAF file is synchronized with GitHub for version control before ELAN is used to annotate and enrich the data, and the human-in-the-loop procedure is used to validate the results with continuous integration and deployment (CI/CD) pipelines providing automated feedback on the annotations. Details on these procedures are provided in section \ref{human-in-the-loop}. 

The validated results are handed over to the EASELAN Post-processing component (Box C in Fig \ref{fig:pipeline}). Here, the results are prepared for the external report and exported into desired formats, such as CSV files, GitHub pages, or knowledge bases, as described in section \ref{post-processing}. Furthermore, EASELAN can be extended to the train Machine Learning models for automatic annotation and dissemination, as well as to export into Narrative Enabled Episodic Memories (NEEMs, see chapter \ref{use case}). 

\subsection{EASELAN Pre-Processing Component}  
\label{pre-processing}
The pre-processing in the EASELAN workflow (Box A in Fig \ref{fig:pipeline}) aims to improve data preparation and consistency across the various data modalities. Multimodal data can be stored locally or accessed remotely from servers. To track changes, we recommend employing data version control systems like LakeFS\footnote{https://lakefs.io/}, which can be seamlessly integrated into the pipeline by modifying file reading and writing scripts. The pre-processing pipeline converts audio and video data based on ELAN recommendations \cite{elan_video_guidelines}, transcribes audio to text, and adapts biosignal data into an efficient format. The pipeline ends with preparing the EAF files containing all data for further use with the ELAN tool.

\subsubsection{Audio and Video Conversion}
The EASELAN workflow maintains compatibility with ELAN and uses FFMPEG\footnote{https://www.ffmpeg.org/} as a free open-source command-line-based tool to encode or convert videos and audio according to the recommended specifications. Thus, we leverage the ELAN features of supporting many file formats, with varying degrees of upward and downward compatibility, depending on available media player frameworks \cite{elan_video_guidelines}. Audio data includes all acoustic events during recording and can provide information about the environment, the acoustic scene, and spoken communications.

\subsubsection{Speech Transcription}
EASELAN allows to integrate any speech recognition engine to provide transcriptions of spoken communication. While we formerly used our inhouse real-time decoder for biosignal processing BioKIT  \cite{telaar2014biokit}, we received more robust speech transcription results with Whisper, OpenAI's multilingual automatic speech recognition system for the transcription and translation of audio and video files \cite{radford2023robust}. To read and transcribe speech, the word-timestamps parameter is set to TRUE for word-level transcription to better align with the input file. The generated output files include the text, word segments, and the detected language. 
Speech recognition is particularly useful in scenarios where humans are asked to comment on their behavior by thinking aloud. Concurrent and retrospective \textit{think alouds} can be leveraged to support the annotation and understanding of everyday activities \cite{meier_comparative_2019}, see also chapter \ref{use case}.

\subsubsection{Biosignal Formatting}
Human activities and internal states such as emotion or attention can be inferred from biosignals such as EMG, ECG and EEG. These biosignals are typically high-dimensional with up to 256 channels for EEG. To make the storage, handling, and analyses of signal data more compatible across platforms and software tools, EASELAN provides a biosignal formatting module to optionally convert these data into CSV (Comma-Separated Values) formatted files. Here, the EASELAN formatting assumes that artifact removal and noise filters were already applied at the time of capturing and storing the respective signal. However, EASELAN allows to integrate any script for artifact removal and de-noising filters, and similar to process data.

\subsubsection{EAF File Preparation}
After finalizing the processing and formatting of multimodal data input, each pre-processed data stream is stored. For this, EASELAN adopts the EUDICO Annotation Format (EAF) and leverages the `tiers' concept of ELAN \cite{elan2004}, which displays annotations in layers. Tiers can be independent or linked, and types of tiers are preset based on constraints. The resulting EAF file stores the information about tiers, annotations, and media files. 

Speech transcriptions from Whisper are included as the ``Whisper Transcripts" tier, with corresponding video and audio files linked to the annotation document. The integration of biosignals is achieved through ELAN's \emph{secondary data} functionality. Secondary data annotations are not directly derived from modalities other than video and audio. The primary challenge in automating secondary data integration lies in its representation. Unlike annotations, secondary data are not stored in EAF files but in additional \texttt{\_tsconf.xml} and \texttt{.pfsx} files. These files typically store Graphical User Interface (GUI) configuration details and represent personal annotator preferences. They are excluded from version control but can be recreated during post-processing. For this purpose, we used the Python package Pympi \cite{pympi-1.70}, which facilitates working with these types of files.

\subsection{EASELAN Human-in-the-Loop Annotation Component}
\label{human-in-the-loop}
The central concept of the EASELAN framework is to support the continuous refinement and extension of annotations. This is achieved by the Human-in-the-loop Annotation component of EASELAN (Box B in Fig \ref{fig:pipeline}). 
Annotations are created by a variety of human annotator experts, organized in a distributed process over time. To facilitate continuous refinement and extension performed by many annotators in a distributed process over time, synchronization and version control is required - this is established by a \textit{Continuous Integration / Continuous Deployment} pipeline. Furthermore, human annotators need a powerful and flexible visualization of all tiers to keep track of the annotation process. For this visualization, the \textit{EASELAN Dashboard} is created, a multitier format similar to a full score in music.  In addition, to support semi-supervised annotation processes, which are facilitated by machine learning-based models, we propose a human-in-the-loop design that also involves an \textit{Annotation Feedback Loop} that executes checks and tests to ensure highest quality standards and consistency across different annotators and models.  

\subsubsection{GitHub and Continuous Integration / Continuous Deployment Pipeline}
To guarantee seamless teamwork and progress tracking among distributed human annotators, EASELAN uses Git repositories. In the EASELAN workflow, EAF files are uploaded to the Git repositories and changes made during annotation are tracked. This process is automated through the Python package \texttt{pygit2\footnote{https://pypi.org/project/pygit2/}}. All relevant Git information, including repository URL, author credentials, and branch name, is updated in a configuration file. This ensures that collaborators can access details about all activities performed within the repository, including data access, modifications, and associated user information.
EASELAN provides a Continuous Integration/Continuous Deployment (CI/CD) pipeline that is regularly synchronized with the Git repository. The CI/CD pipeline is an iterative process that automates validation and testing, including verification and spell-checking of annotations against predefined templates. Feedback is provided in HTML format via GitHub Pages. This step aims to streamline the general annotation process and to support the reliability of annotated data while keeping annotators and project members informed of changes and progress.

\begin{figure}[H]
\centering
\includegraphics[width=1.0\linewidth]{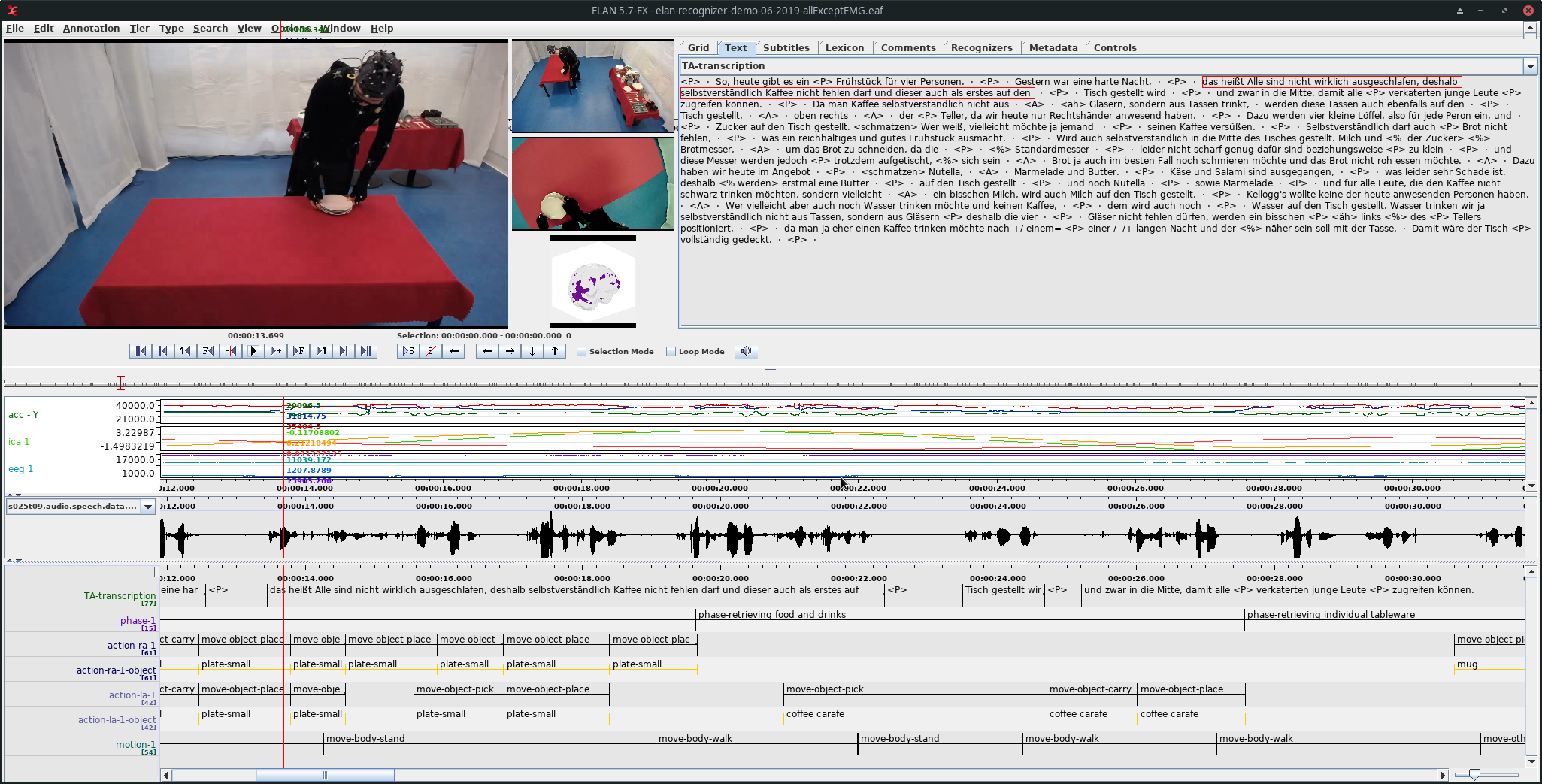}
\caption{\label{fig:biosignals}Screenshort of the EASELAN Dashboard (description see subsection \ref{dashboard}). 
}
\end{figure}

\subsubsection{EASELAN Dashboard}
\label{dashboard}
The EASELAN dashboard is designed to provide the best possible support to humans in annotating high-dimensional, complex, multimodal data. It has a format similar to a full score in music, where all tiers are arranged in a partitur. Fig \ref{fig:biosignals} shows an example of the EASELAN Dashboard built on the ELAN tool. Here, the upper left corner shows a video stream with three selected camera perspectives, the upper right shows WHISPER-transcribed think alouds that were produced by the study participant shown in the camera. The participants carry out a Table Setting Task (see also section \ref{use case}). The lower panel shows the time aligned biosignals, from top to bottom: (a) functional MRI brain activity of an individual, observing the human activity carried out by the participant, (b) acceleration data of the participant' body motion, (c) Independent Component Analysis of EEG for artifact removal, (d) raw EEG, speech signal of think aloud, (e) action annotations of different levels, object annotations, and motion recognition. The red bar shows the current position in which the multimodal signal data is displayed synchronously.

\subsubsection{Annotation Feedback Loop}
Integration of annotation tools with version control systems such as Git facilitates seamless synchronization and collaboration, while CI/CD pipelines automate validation and testing of annotated files, enhancing efficiency and reliability. EASELAN employs an iterative process in which human annotations synchronize with the Git repository. The core concept is to facilitate the continuous improvement of annotations while keeping users informed.
For this purpose, EAF files are accessed from the Git repository. After annotation in ELAN using the EASELAN Dashboard, the files are pushed to the repository, triggering the CI/CD pipeline. With YAML\footnote{https://yaml.org/}, a human-readable data serialization language, a configuration file is created that defines pipeline stages and jobs, including spell checking and annotation verification. CI/CD runners then execute jobs concurrently in the specified order. The job status and details are displayed in the CI/CD pipeline.
In the EASELAN workflow, the test stage includes the spell-checking of transcripts and the verification of annotations against predefined templates. It also checks for empty tiers and ensures compatibility with controlled vocabularies. Feedback is provided in HTML format via the GitHub pages, and annotations are uploaded with executed tests. This entire workflow ensures data quality early while streamlining annotation, promoting collaboration, and improving reliability.

\subsection{EASELAN Post-processing Component}
\label{post-processing}
The final preparation steps for data preservation and release are performed in the EASELAN Post-processing component (Box C in Fig \ref{fig:pipeline}). Similarly to our argument for dashboards, we believe that good visualization helps maintain an overview and minimize errors. Therefore, the EASELAN post-processing workflow applies data visualization to interpret the resulting data. It automatically generates comprehensive annotation reports and uses GitHub pages and static websites from the repository, which are deployed by the CI/CD pipeline. The reports include spell-check and annotation verification reports. In addition, there is an option to download these reports as a CSV file for further analysis.

\section{EASELAN Use Case: The EASE Table Setting Database}
\label{use case}
The EASELAN use case briefly described in this section originates from a recently completed EASE Table Setting Database (EASE-TSD) project of the Cognitive Systems Lab (CSL) at the University of Bremen \cite{mason2018TSD,Schultz-TSD2025} and provides a practical example of using the EASELAN framework for the entire data management process from data acquisition to data preservation and dissemination. 

The EASE-TSD was collected, processed and disseminated in the context of the DFG Collaborative Research Center ``Everyday Activities Science and Engineering (EASE)"\footnote{http://ease-crc.org}. EASE advances the understanding of how human-scale manipulation tasks can be mastered by robotic agents. In particular, EASE equips cognitive robots with human-like reasoning, learned from fine-grained multimodal observations of open-world human activities. This requires a robust processing pipeline for holistic observation of human behavior \cite{meier2018synchronized}, which integrates multimodal data collection, biosignal processing, hierarchical annotations, and ontological reasoning of human activities\cite{MasonEtAl2021_FromHumantoRobot}. 

\subsection{EASE Table Setting Dataset (EASE-TSD)}
EASE-TSD is a dataset of multimodal high-dimensional biosignals recorded synchronously from human subjects who are setting a table (breakfast or dinner for 2 or 4 persons in formal or informal style) in a controlled laboratory setup. 100 participants were recorded and data from 78 sessions are available, each recorded during six table-setting trials using eight synchronized biosignal streams, which capture with 22 sensors for both the planning and execution of human behavior: marker-based motion capturing, environmental and first-person video cameras (7 RGB cameras), eye-tracking, 16-channel EMG on both hands, electrodermal activity, acceleration, near- and farfield microphones (2 channel), and 16-channel EEG. All biosignals were synchronized and recorded using the Lab Streaming Layer (LSL) \cite{kothe2025lab}. In total, EASE-TSD has over 430 hours of sensor data. 

EASE-TSD data was captured using distributed recording software, which stored the data on different hard drives. EASELAN uses a central backed-up storage, so for EASE-TSD this was implemented with a version-managed data lake that allows the storage of raw and processed data. Depending on the purpose of usage, the annotation files were exported into different formats (see below). 

\subsection{EASE-TSD Human-in-the-loop Annotation}
Participants were instructed to think aloud concurrently and retrospectively to explain and comment on their table-setting actions and the corresponding cognitive processes \cite{meier_comparative_2019}. 

After recording, the EASE-TSD data we subjected to semi-automatic labeling, post-processing, and analysis procedures, utilizing the latest biosignal processing and machine learning methods. Human annotators used the EASELAN workflow for efficient and consistent 3-level annotation including phase, action, and motion levels, further divided by hands, head, and body. An additional tier contains objects for each hand (e.g., cups, plates, cutlery). Here, we used EASELAN with a GitLab-Issue system for task prioritization to support and speed-up the annotation process. 

The speech recordings of the think-aloud (TA) protocols, in which participants verbalized actions at different abstraction levels during and after the task, were transcribed using Whisper. The Whisper transcription files were stored in ELAN files, and manually cross-checked by the human annotators. In addition,  TA protocols were annotated using TA codes. 

In total, more than 260 hours of EASE-TSD data are annotated. 

\subsection{EASE-TSD Post-processing and Dissemination}
In post-processing, annotation tiers and biosignals we merged into a single file per run, which served as a partitur file for a structured dataset overview. 

\begin{figure}[H]
\centering
\includegraphics[width=1.0\linewidth]{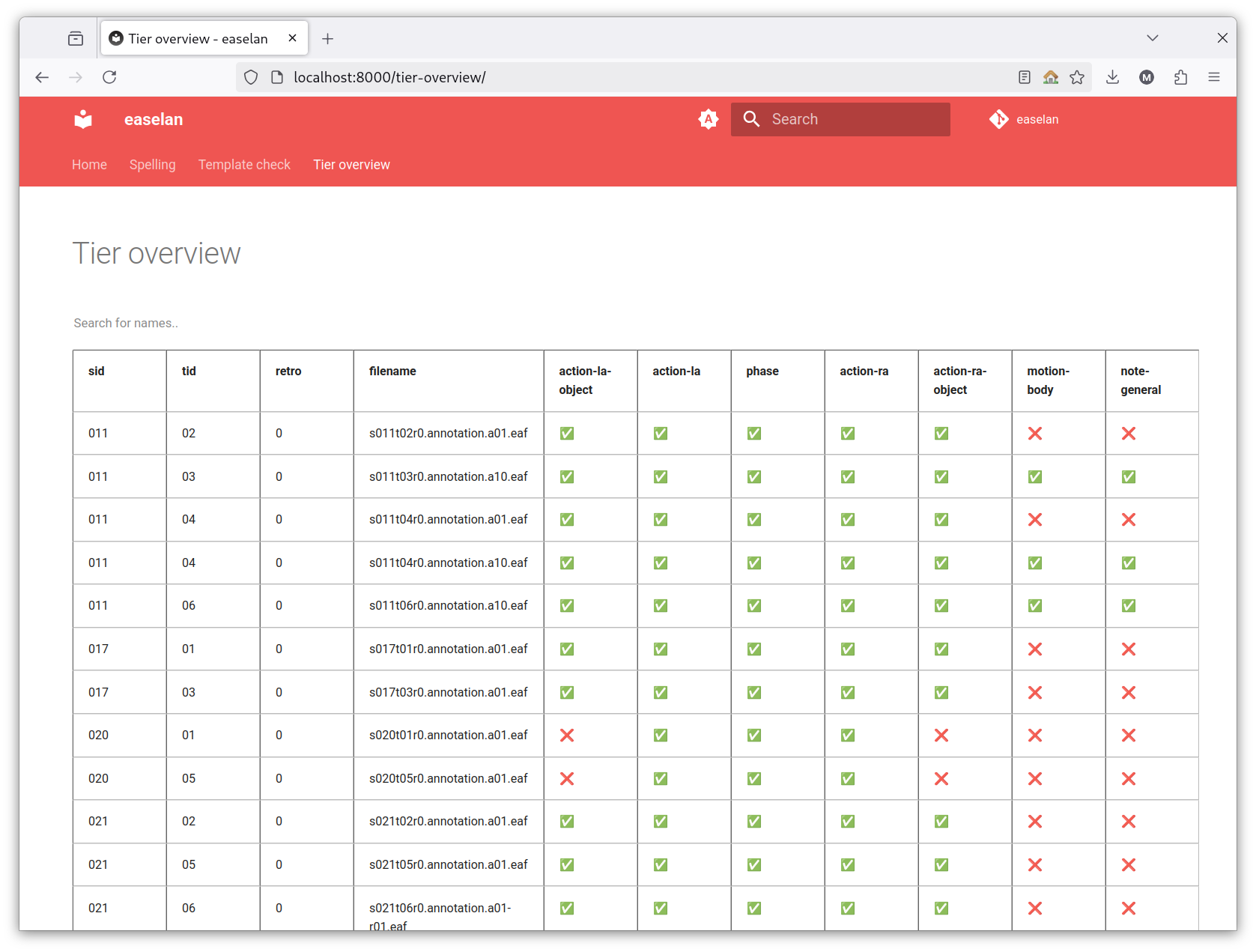}
\caption{\label{fig:tier}Tier overview in HTML-export for validation and testing. Check-marks or red 'X' indicate if a tier is present in an annotation or not present/contains no labels.}
\end{figure}

For verification and testing, the EAF annotation files (with suffix \texttt{.eaf}) were checked against predefined ELAN templates to identify missing tiers and labels. Also, the transcripts were spell-checked. The verification scripts produce a pandas data frame with information on each EAF file which can be easily exported as Markdown. For simplicity reasons, the static side generator \texttt{mkdocs} is used, creating a cohesive website. Figure \ref{fig:tier} illustrates the annotation verification process in HTML format. The workflow, various data overviews, and the HTML overview provided automatic feedback to the annotators and helped to identify any necessary adjustments. The carefully validated and finalized EASE-TSD was provided for research purposes via the Open Science Framework \cite{HartmannMeierPutzeSchultz2025}. 

\subsection{From Human to Robot Everyday Activity}
The EASE-TSD was transferred to the openEASE cloud-based framework for knowledge representation and reasoning \cite{Beetz_Tenorth_Winkler_2015}, to facilitate joint research in the EASE Collaborative Research Center \cite{MasonEtAl2021_FromHumantoRobot}. In openEASE, the richly annotated holistic information on human activities based on multimodal data and biosignals is accessible to cognitive robots via Narrative-enabled Episodic Memories (NEEMs). NEEMS are structured recordings of actions performed by either a robot or a human operator, captured in real-world, simulated or virtual reality environments. NEEMs include both sub-symbolic low-level data and symbolic semantically annotated information. The sharing of NEEMs between humans and robots takes place via an ontology that is shared across all NEEMs. This is facilitated by the Socio-physical Model of Activities (SOMA) ontology, an ontological modeling approach for autonomous robotic agents performing everyday manipulation tasks, developed within EASE \cite{beßler2020foundationssociophysicalmodelactivities}. Since SOMA is shared across all NEEMs, it allows to align and compare NEEMs - even when recorded using different tools or under varying conditions\footnote{https://vrb.ease-crc.org/explore-labs/neem-lab/}. 

The SOMA ontology provides a semantic model to represent the locations of objects, regions of interest, and agents' movements. It can be used for robotic, human, and virtual agent activities alike \cite{kunze2018soma}. The SOMA concepts are applied to automatically generate a controlled vocabulary in the ELAN (\texttt{.ecv} file) which in turn can be used in the EASELAN annotation. Once the annotation files have been pushed to the git repository, they are picked up by an export script to generate an semantic representation using the \texttt{.rfd} file format. A semantic database (like Apache Jena or OpenEASE) or a simple RDF-library (e.g. rdflib) can then be used to post semantic queries in order to yield data for further analysis or model building. This concept was leveraged to search and retrieve EASE-TSD data subsets to create 3D-object tracking \cite{bredereke2025modularpipeline3dobject}, gait parameter estimation \cite{hartmannkloeckner} and informed models for attention detection from EEG in human-robot \cite{Richter_Putze_Ivucic_Brandt_Schütze_Reisenhofer_Wrede_Schultz_2023}. 

As a result, the EASE-TSD human activities database along with the data analysis and annotation pipeline EASELAN provides a rich ground truth and foundation to advance cognition-enhanced robots for everyday activities \cite{MasonEtAl2021_FromHumantoRobot}.

\section{Expected Impact of Frameworks like EASELAN}
Annotating complex and heterogeneous datasets with multimodal biosignals is a difficult and time-consuming task. The open-source EASELAN framework aids researchers and developers by providing optimized workflows and pipelines. EASELAN supports synchronization and provides visualization of biosignals within the annotation structure of ELAN and includes CI/CD pipelines for validation, spell checking, and reviewing annotation quality. EASELAN tracks all changes to annotations and datasets via Git-based version control and thus facilitates shared annotation processes over time. Likewise, EASELAN's automated processing and annotation pipelines provide more consistent formatting to help standardize workflows. Together, these features are designed to promote data management, access, and collaboration across distributed laboratories and disciplinary boundaries.

Given the enormous progress in machine learning and the great potential for innovative biosignal-adaptive systems that fuse multimodal data, EASELAN closes an important gap in understanding and explaining the impact of biosignals for the holistic interpretation of human behavior. Instead of simply extracting more features from increasingly powerful multimodal datasets, a unified visual annotation environment such as EASELAN enables a holistic, contextualized annotation of complex data structures. Our software is located at a critical junction, where interdisciplinary researchers can advance their understanding of a wide range of human behaviors and signals by easily visualizing and handling even extremely complex multimodal data including high-dimensional biosignals.

\section{Summary and Conclusion}
We introduced EASELAN, an open-source framework for multimodal biosignal annotation and data management. It is a comprehensive human-in-the-loop framework for annotating high-dimensional sensor data. It aligns multiple modalities, including audio, text, video, and biosignals resulting from human behavior and their activities of the eyes, body, skin, muscles, and brain. EASELAN builds upon the well-established ELAN tool to streamline integration, synchronization, formatting, processing, transcription, annotation, access, sharing, and preservation of multimodal data. It leverages Git for version control and Continuous Integration/Continuous Deployment pipelines for annotation validation and automated user feedback. 

Although EASELAN provides multiple advancements, some limitations remain. The integration of more accurate AI-assisted event detection could further improve annotation consistency and further reduce the workload of human annotators. Additionally, the machine learning-based pipeline to continuously train and validate event detection models is still limited. Here, future work could include direct export options for frameworks such as TensorFlow or PyTorch, as well as better compatibility with external databases. 

We conclude that EASELAN facilitates a more structured and reproducible approach to data management in a familiar and time-tested environment. As research in biosignal-adaptive AI systems and multimodal fusion continues to advance, improving the community's ability to reproducibly annotate and manage highly complex multimodal data sets will be crucial to extend our understanding of the role of biosignals in a broad range of contexts. In this paper, we argue that advances in annotation software will have an important role to play, for example, when embracing interdisciplinary expertise and establishing ground truths. EASELAN is open-source and readily available for interdisciplinary contributions to further extend these capabilities \cite{MeierRammohanSchultz2025}.

\section*{Acknowledgements}
\label{}
\appendix
\label{sec:sample:appendix}

This research was partially funded by the German Research Foundation DFG, as part of Collaborative Research Center (Sonderforschungsbereich) 1320 Project-ID 329551904 “EASE - Everyday Activity Science and Engineering”, University of Bremen, Germany (http://www.ease-crc.org/).


 \bibliographystyle{elsarticle-num} 
 \bibliography{References}

\begin{thebibliography}{10}
\expandafter\ifx\csname url\endcsname\relax
  \def\url#1{\texttt{#1}}\fi
\expandafter\ifx\csname urlprefix\endcsname\relax\def\urlprefix{URL }\fi
\expandafter\ifx\csname href\endcsname\relax
  \def\href#1#2{#2} \def\path#1{#1}\fi

\bibitem{HartmannMeierPutzeSchultz2025}
{Hartmann, Y and Meier, M and Putze, F and Schultz T}, \href{https://osf.io/rbyfk/?view_only=72d82e6a1b1a4ae9b09d67e6837e5412}{{EASE-TSD OSF Data Repository}}, accessed: 2025-10-05 (n.d.).
\newline\urlprefix\url{https://osf.io/rbyfk/?view_only=72d82e6a1b1a4ae9b09d67e6837e5412}

\bibitem{schultz_biosignals_2023}
T.~Schultz, A.~Maedche, \href{https://doi.org/10.1007/s42452-023-05412-w}{Biosignals meet {A}daptive {S}ystems}, {SN} Applied Sciences 5~(9) (2023) 234.
\newblock \href {https://doi.org/10.1007/s42452-023-05412-w} {\path{doi:10.1007/s42452-023-05412-w}}.
\newline\urlprefix\url{https://doi.org/10.1007/s42452-023-05412-w}

\bibitem{wittenburg-etal-2006-elan}
P.~Wittenburg, H.~Brugman, A.~Russel, A.~Klassmann, H.~Sloetjes, \href{http://www.lrec-conf.org/proceedings/lrec2006/pdf/153_pdf.pdf}{{ELAN}: a {P}rofessional {F}ramework for {M}ultimodality {R}esearch}, in: Proceedings of the Fifth International Conference on Language Resources and Evaluation ({LREC}{'}06), European Language Resources Association (ELRA), Genoa, Italy, 2006, pp. 1556--1559.
\newline\urlprefix\url{http://www.lrec-conf.org/proceedings/lrec2006/pdf/153_pdf.pdf}

\bibitem{kipp_anvil_2001}
M.~Kipp, \href{https://www.isca-speech.org/archive/eurospeech_2001/kipp01_eurospeech.html}{{ANVIL} - {A} {G}eneric {A}nnotation {T}ool for {M}ultimodal {D}ialogue}, in: 7th European Conference on Speech Communication and Technology (Eurospeech 2001), {ISCA}, 2001, pp. 1367--1370.
\newblock \href {https://doi.org/10.21437/Eurospeech.2001-354} {\path{doi:10.21437/Eurospeech.2001-354}}.
\newline\urlprefix\url{https://www.isca-speech.org/archive/eurospeech_2001/kipp01_eurospeech.html}

\bibitem{naghib_comprehensive_2023}
A.~Naghib, N.~Jafari~Navimipour, M.~Hosseinzadeh, A.~Sharifi, \href{https://doi.org/10.1007/s11276-022-03177-5}{A {C}omprehensive and {S}ystematic {L}iterature {R}eview on the {B}ig {D}ata {M}anagement {T}echniques in the {I}nternet of {T}hings}, Wireless Networks 29~(3) (2023) 1085--1144.
\newblock \href {https://doi.org/10.1007/s11276-022-03177-5} {\path{doi:10.1007/s11276-022-03177-5}}.
\newline\urlprefix\url{https://doi.org/10.1007/s11276-022-03177-5}

\bibitem{schultz_lablinking_2024}
T.~Schultz, F.~Putze, R.~Reisenhofer, T.~Fehr, M.~Meier, C.~Mason, F.~Ahrens, \href{https://doi.org/10.1007/s42452-024-06122-7}{{LabLinking}: {T}heory, {F}ramework, and {S}olutions of {C}onnecting {L}aboratories for {D}istributed {H}uman {E}xperiments}, Discover Applied Sciences 6~(9) (2024) 448.
\newblock \href {https://doi.org/10.1007/s42452-024-06122-7} {\path{doi:10.1007/s42452-024-06122-7}}.
\newline\urlprefix\url{https://doi.org/10.1007/s42452-024-06122-7}

\bibitem{kothe2025lab}
C.~Kothe, S.~Y. Shirazi, T.~Stenner, D.~Medine, C.~Boulay, M.~I. Grivich, F.~Artoni, T.~Mullen, A.~Delorme, S.~Makeig, The {L}ab {S}treaming {L}ayer for {S}ynchronized {M}ultimodal {R}ecording, Imaging Neuroscience 3 (2025) IMAG--a.

\bibitem{MeierRammohanSchultz2025}
{Meier, Moritz}, \href{https://github.com/cognitive-systems-lab/easelan}{{EASELAN} repository}, accessed: 2025-10-05 (n.d.).
\newline\urlprefix\url{https://github.com/cognitive-systems-lab/easelan}

\bibitem{zafar_reviewing_2023}
I.~Zafar, S.~Anwar, F.~kanwal, W.~Yousaf, F.~Un~Nisa, T.~Kausar, Q.~ul~Ain, A.~Unar, M.~A. Kamal, S.~Rashid, K.~A. Khan, R.~Sharma, \href{https://www.sciencedirect.com/science/article/pii/S1746809423006961}{Reviewing {M}ethods of {D}eep {L}earning for {I}ntelligent {H}ealthcare {S}ystems in {G}enomics and {B}iomedicine}, Biomedical Signal Processing and Control 86 (2023) 105263.
\newblock \href {https://doi.org/10.1016/j.bspc.2023.105263} {\path{doi:10.1016/j.bspc.2023.105263}}.
\newline\urlprefix\url{https://www.sciencedirect.com/science/article/pii/S1746809423006961}

\bibitem{liu2023largelanguagemodelsfewshot}
X.~Liu, D.~McDuff, G.~Kovacs, I.~Galatzer-Levy, J.~Sunshine, J.~Zhan, M.-Z. Poh, S.~Liao, P.~D. Achille, S.~Patel, \href{https://arxiv.org/abs/2305.15525}{Large {L}anguage {M}odels are {F}ew-{S}hot {H}ealth {L}earners} (2023).
\newblock \href {http://arxiv.org/abs/2305.15525} {\path{arXiv:2305.15525}}.
\newline\urlprefix\url{https://arxiv.org/abs/2305.15525}

\bibitem{steinert_evaluation_2022}
L.~Steinert, F.~L. Kölling, F.~Putze, D.~Küster, T.~Schultz, {E}valuation of an {E}ngagement-{A}ware {R}ecommender {S}ystem for {P}eople with {D}ementia, in: Proceedings of the 30th {ACM} Conference on User Modeling, Adaptation and Personalization, 2022, pp. 89--98.

\bibitem{steinert_predicting_2022}
L.~Steinert, F.~Putze, D.~Küster, T.~Schultz, Predicting {A}ctivation {L}iking of {P}eople with {D}ementia, Frontiers in Computer Science 3 (2022) 770492.

\bibitem{obaid_endowing_2018}
M.~Obaid, R.~Aylett, W.~Barendregt, C.~Basedow, L.~J. Corrigan, L.~Hall, A.~Jones, A.~Kappas, D.~Küster, A.~Paiva, Endowing a {R}obotic {T}utor with {E}mpathic {Q}ualities: {D}esign and {P}ilot {E}valuation, International Journal of Humanoid Robotics 15~(6) (2018) 1850025, publisher: World Scientific.

\bibitem{meier_comparative_2019}
M.~Meier, C.~Mason, F.~Putze, T.~Schultz, \href{https://www.isca-speech.org/archive/interspeech_2019/meier19_interspeech.html}{Comparative {A}nalysis of {T}hink-{A}loud {M}ethods for {E}veryday {A}ctivities in the {C}ontext of {C}ognitive {R}obotics}, in: Interspeech 2019, {ISCA}, 2019, pp. 559--563.
\newblock \href {https://doi.org/10.21437/Interspeech.2019-3072} {\path{doi:10.21437/Interspeech.2019-3072}}.
\newline\urlprefix\url{https://www.isca-speech.org/archive/interspeech_2019/meier19_interspeech.html}

\bibitem{muhammad_comprehensive_2021}
G.~Muhammad, F.~Alshehri, F.~Karray, A.~E. Saddik, M.~Alsulaiman, T.~H. Falk, \href{https://www.sciencedirect.com/science/article/pii/S1566253521001330}{A {C}omprehensive {S}urvey on {M}ultimodal {M}edical {S}ignals {F}usion for {S}mart {H}ealthcare {S}ystems}, Information Fusion 76 (2021) 355--375.
\newblock \href {https://doi.org/10.1016/j.inffus.2021.06.007} {\path{doi:10.1016/j.inffus.2021.06.007}}.
\newline\urlprefix\url{https://www.sciencedirect.com/science/article/pii/S1566253521001330}

\bibitem{dablain_deepsmote_2023}
D.~Dablain, B.~Krawczyk, N.~V. Chawla, \href{https://ieeexplore.ieee.org/abstract/document/9694621}{{DeepSMOTE}: {Fusing} {Deep} {Learning} and {SMOTE} for {Imbalanced} {Data}}, IEEE Transactions on Neural Networks and Learning Systems 34~(9) (2023) 6390--6404, conference Name: IEEE Transactions on Neural Networks and Learning Systems.
\newblock \href {https://doi.org/10.1109/TNNLS.2021.3136503} {\path{doi:10.1109/TNNLS.2021.3136503}}.
\newline\urlprefix\url{https://ieeexplore.ieee.org/abstract/document/9694621}

\bibitem{fernandez_smote_2018}
A.~Fernandez, S.~Garcia, F.~Herrera, N.~V. Chawla, \href{https://www.jair.org/index.php/jair/article/view/11192}{{SMOTE} for {Learning} from {Imbalanced} {Data}: {Progress} and {Challenges}, {Marking} the 15-year {Anniversary}}, Journal of Artificial Intelligence Research 61 (2018) 863--905.
\newblock \href {https://doi.org/10.1613/jair.1.11192} {\path{doi:10.1613/jair.1.11192}}.
\newline\urlprefix\url{https://www.jair.org/index.php/jair/article/view/11192}

\bibitem{lee_artificial_2024}
Y.~J. Lee, C.~Park, H.~Kim, S.~J. Cho, W.-H. Yeo, \href{https://doi.org/10.1007/s44258-024-00043-1}{Artificial {I}ntelligence on {B}iomedical {S}ignals: {T}echnologies, {A}pplications, and {F}uture {D}irections}, Med-X 2~(1) (2024) 25.
\newblock \href {https://doi.org/10.1007/s44258-024-00043-1} {\path{doi:10.1007/s44258-024-00043-1}}.
\newline\urlprefix\url{https://doi.org/10.1007/s44258-024-00043-1}

\bibitem{derdiyok_biosignal_2023}
S.~Derdiyok, F.~P. Akbulut, \href{https://doi.org/10.1007/s00530-023-01071-4}{Biosignal {b}ased {E}motion-{O}riented {V}ideo {S}ummarization}, Multimedia Systems 29~(3) (2023) 1513--1526.
\newblock \href {https://doi.org/10.1007/s00530-023-01071-4} {\path{doi:10.1007/s00530-023-01071-4}}.
\newline\urlprefix\url{https://doi.org/10.1007/s00530-023-01071-4}

\bibitem{Barandasetal}
M.~Barandas, D.~Folgado, L.~Fernandes, S.~Santos, M.~Abreu, P.~Bota, H.~Liu, T.~Schultz, H.~Gamboa, \href{https://www.sciencedirect.com/science/article/pii/S2352711020300017}{{TSFEL}: {T}ime {S}eries {F}eature {E}xtraction {L}ibrary}, SoftwareX 11 (2020) 100456.
\newblock \href {https://doi.org/https://doi.org/10.1016/j.softx.2020.100456} {\path{doi:https://doi.org/10.1016/j.softx.2020.100456}}.
\newline\urlprefix\url{https://www.sciencedirect.com/science/article/pii/S2352711020300017}

\bibitem{bian_understanding_2024}
Y.~Bian, D.~Küster, H.~Liu, E.~G. Krumhuber, \href{https://www.mdpi.com/1424-8220/24/1/126}{{U}nderstanding {N}aturalistic {F}acial {E}xpressions with {D}eep {L}earning and {M}ultimodal {L}arge {L}anguage {M}odels}, Sensors 24~(1) (2024) 126, number: 1 Publisher: Multidisciplinary Digital Publishing Institute.
\newblock \href {https://doi.org/10.3390/s24010126} {\path{doi:10.3390/s24010126}}.
\newline\urlprefix\url{https://www.mdpi.com/1424-8220/24/1/126}

\bibitem{salous_smarthelm_2022}
M.~Salous, D.~Küster, K.~Scheck, A.~Dikfidan, T.~Neumann, F.~Putze, T.~Schultz, \href{https://ieeexplore.ieee.org/abstract/document/9945155}{{SmartHelm}: User {S}tudies from {L}ab to {F}ield for {A}ttention {M}odeling}, in: 2022 {IEEE} International Conference on Systems, Man, and Cybernetics ({SMC}), 2022, pp. 1012--1019, {ISSN}: 2577-1655.
\newblock \href {https://doi.org/10.1109/SMC53654.2022.9945155} {\path{doi:10.1109/SMC53654.2022.9945155}}.
\newline\urlprefix\url{https://ieeexplore.ieee.org/abstract/document/9945155}

\bibitem{bodaghi_multimodal_2024}
M.~Bodaghi, M.~Hosseini, R.~Gottumukkala, \href{https://ieeexplore.ieee.org/document/10586177}{A {M}ultimodal {I}ntermediate {F}usion {N}etwork with {M}anifold {L}earning for {S}tress {D}etection}, in: 2024 {IEEE} 3rd International Conference on Computing and Machine Intelligence ({ICMI}), 2024, pp. 1--8.
\newblock \href {https://doi.org/10.1109/ICMI60790.2024.10586177} {\path{doi:10.1109/ICMI60790.2024.10586177}}.
\newline\urlprefix\url{https://ieeexplore.ieee.org/document/10586177}

\bibitem{quiles_perez_data_2024}
M.~Quiles~Pérez, E.~T. Martínez~Beltrán, S.~López~Bernal, E.~Horna~Prat, L.~Montesano Del~Campo, L.~Fernández~Maimó, A.~Huertas~Celdrán, \href{https://www.sciencedirect.com/science/article/pii/S1566253524000095}{{D}ata {F}usion in {N}euromarketing: {M}ultimodal {A}nalysis of {B}iosignals, {L}ifecycle {S}tages, {C}urrent {A}dvances, {D}atasets, {T}rends, and {C}hallenges}, Information Fusion 105 (2024) 102231.
\newblock \href {https://doi.org/10.1016/j.inffus.2024.102231} {\path{doi:10.1016/j.inffus.2024.102231}}.
\newline\urlprefix\url{https://www.sciencedirect.com/science/article/pii/S1566253524000095}

\bibitem{steinert_audio-visual_2021}
L.~Steinert, F.~Putze, D.~Küster, T.~Schultz, \href{https://www.isca-speech.org/archive/interspeech_2021/steinert21_interspeech.html}{Audio-{Visual} {Recognition} of {Emotional} {Engagement} of {People} with {Dementia}}, in: Interspeech 2021, ISCA, 2021, pp. 1024--1028.
\newblock \href {https://doi.org/10.21437/Interspeech.2021-567} {\path{doi:10.21437/Interspeech.2021-567}}.
\newline\urlprefix\url{https://www.isca-speech.org/archive/interspeech_2021/steinert21_interspeech.html}

\bibitem{olugbade_human_2022}
T.~Olugbade, M.~Bieńkiewicz, G.~Barbareschi, V.~D’amato, L.~Oneto, A.~Camurri, C.~Holloway, M.~Björkman, P.~Keller, M.~Clayton, A.~C. D.~C. Williams, N.~Gold, C.~Becchio, B.~Bardy, N.~Bianchi-Berthouze, \href{https://dl.acm.org/doi/10.1145/3534970}{Human {M}ovement {D}atasets: {A}n {I}nterdisciplinary {S}coping {R}eview}, {ACM} Comput. Surv. 55~(6) (2022) 126:1--126:29.
\newblock \href {https://doi.org/10.1145/3534970} {\path{doi:10.1145/3534970}}.
\newline\urlprefix\url{https://dl.acm.org/doi/10.1145/3534970}

\bibitem{via2019}
A.~Dutta, A.~Zisserman, \href{https://dl.acm.org/doi/10.1145/3343031.3350535}{The {VIA} {A}nnotation {S}oftware for {I}mages, {A}udio and {V}ideo}, in: Proceedings of the 27th ACM International Conference on Multimedia, MM ’19, Association for Computing Machinery, New York, NY, USA, 2019, p. 2276–2279.
\newblock \href {https://doi.org/10.1145/3343031.3350535} {\path{doi:10.1145/3343031.3350535}}.
\newline\urlprefix\url{https://dl.acm.org/doi/10.1145/3343031.3350535}

\bibitem{webanno2013}
S.~M. Yimam, I.~Gurevych, R.~Eckart~de Castilho, C.~Biemann, \href{https://aclanthology.org/P13-4001}{{WebAnno}: A {F}lexible, {W}eb-based and {V}isually {S}upported {S}ystem for {D}istributed {A}nnotations}, in: M.~Butt, S.~Hussain (Eds.), Proceedings of the 51st Annual Meeting of the Association for Computational Linguistics: System Demonstrations, Association for Computational Linguistics, Sofia, Bulgaria, 2013, p. 1–6.
\newline\urlprefix\url{https://aclanthology.org/P13-4001}

\bibitem{cvat_tool}
OpenCV, Intel, \href{https://docs.cvat.ai/docs/}{Computer Vision Annotation Tool (CVAT) Documentation}, accessed: 2024-12-01 (2024).
\newline\urlprefix\url{https://docs.cvat.ai/docs/}

\bibitem{friard_boris_2016}
O.~Friard, M.~Gamba, \href{https://onlinelibrary.wiley.com/doi/abs/10.1111/2041-210X.12584}{{BORIS}: {A} {F}ree, {V}ersatile {O}pen-source {E}vent-logging {S}oftware for {V}ideo/{A}udio {C}oding and {L}ive {O}bservations}, Methods in Ecology and Evolution 7~(11) (2016) 1325--1330.
\newblock \href {https://doi.org/10.1111/2041-210X.12584} {\path{doi:10.1111/2041-210X.12584}}.
\newline\urlprefix\url{https://onlinelibrary.wiley.com/doi/abs/10.1111/2041-210X.12584}

\bibitem{anvil2012}
M.~Kipp, \href{https://onlinelibrary.wiley.com/doi/abs/10.1002/9781118219546.ch21}{Multimedia Annotation, Querying, and Analysis in Anvil}, John Wiley \& Sons, Ltd, 2012, Ch.~21, p. 351–367.
\newblock \href {https://doi.org/10.1002/9781118219546.ch21} {\path{doi:10.1002/9781118219546.ch21}}.
\newline\urlprefix\url{https://onlinelibrary.wiley.com/doi/abs/10.1002/9781118219546.ch21}

\bibitem{elan2004}
H.~Brugman, A.~Russel, X.~Nijmegen, Annotating {M}ulti-media/{M}ulti-modal {R}esources with {ELAN}., in: LREC, Lisbon, 2004, pp. 2065--2068.

\bibitem{elan_video_guidelines}
{Max Planck Institute for Psycholinguistics}, \href{https://www.mpi.nl/tools/elan/docs/Video_encoding_guidelines_ELAN.pdf}{{V}ideo {E}ncoding {G}uidelines for {ELAN}}, accessed: 2025-02-25 (n.d.).
\newline\urlprefix\url{https://www.mpi.nl/tools/elan/docs/Video_encoding_guidelines_ELAN.pdf}

\bibitem{telaar2014biokit}
D.~Telaar, M.~Wand, D.~Gehrig, F.~Putze, C.~Amma, D.~Heger, N.~T. Vu, M.~Erhardt, T.~Schlippe, M.~Janke, C.~Herff, T.~Schultz, \href{https://www.csl.uni-bremen.de/cms/images/documents/publications/TelaarEtAl_IS14_BioKIT.pdf}{{BioKIT - Real-time Decoder For Biosignal Processing}}, in: The 15th Annual Conference of the International Speech Communication Association, Singapore, 2014, pp. 2650--2654, interspeech 2014.
\newline\urlprefix\url{https://www.csl.uni-bremen.de/cms/images/documents/publications/TelaarEtAl_IS14_BioKIT.pdf}

\bibitem{radford2023robust}
A.~Radford, J.~W. Kim, T.~Xu, G.~Brockman, C.~McLeavey, I.~Sutskever, Robust {S}peech {R}ecognition via {L}arge-scale {W}eak {S}upervision, in: International Conference on Machine Learning, PMLR, 2023, pp. 28492--28518.

\bibitem{pympi-1.70}
M.~Lubbers, F.~Torreira, pympi-ling: a {Python} module for processing {ELAN}s {EAF} and {Praat}s {TextGrid} annotation files., \url{https://pypi.python.org/pypi/pympi-ling}, version 1.70 (2013-2021).

\bibitem{mason2018TSD}
C.~Mason, M.~Meier, F.~Ahrens, T.~Fehr, M.~Herrmann, F.~Putze, T.~Schultz, {Human Activities Data Collection and Labeling using a Think-Aloud Protocol in a Table Setting Scenario}, in: IROS 2018: Workshop on Latest Advances in Big Activity Data Sources for Robotics \& New Challenges, Madrid, Spain, IROS, 2018.

\bibitem{Schultz-TSD2025}
{Meier, M and Hartmann, Y and El El Ouahabi, Y and Bredereke L, and Putze F, and Schultz T}, {Everyday Science and Engineering Table Setting Dataset}, submitted to Applied Sciences, Springer Nature (n.d.).

\bibitem{meier2018synchronized}
M.~Meier, C.~Mason, R.~Porzel, F.~Putze, T.~Schultz, {Synchronized Multimodal Recording of a Table Setting Dataset}, in: Proceedings of the IROS 2018 Workshop on Latest Advances in Big Activity Data Sources for Robotics \& New Challenges (Madrid), 2018.

\bibitem{MasonEtAl2021_FromHumantoRobot}
C.~Mason, K.~Gadzicki, M.~Meier, F.~Ahrens, T.~Kluss, J.~Maldonado, F.~Putze, T.~Fehr, C.~Zetzsche, M.~Herrmann, K.~Schill, T.~Schultz, {From Human to Robot Everyday Activity}, in: 2020 IEEE/RSJ International Conference on Intelligent Robots and Systems (IROS), 2020, pp. 8997--9004.
\newblock \href {https://doi.org/10.1109/IROS45743.2020.9340706} {\path{doi:10.1109/IROS45743.2020.9340706}}.

\bibitem{Beetz_Tenorth_Winkler_2015}
M.~Beetz, M.~Tenorth, J.~Winkler, \href{https://ieeexplore.ieee.org/abstract/document/7139458}{{Open-EASE}}, in: 2015 IEEE International Conference on Robotics and Automation (ICRA), 2015, p. 1983–1990.
\newblock \href {https://doi.org/10.1109/ICRA.2015.7139458} {\path{doi:10.1109/ICRA.2015.7139458}}.
\newline\urlprefix\url{https://ieeexplore.ieee.org/abstract/document/7139458}

\bibitem{beßler2020foundationssociophysicalmodelactivities}
D.~Beßler, R.~Porzel, M.~Pomarlan, A.~Vyas, S.~Höffner, M.~Beetz, R.~Malaka, J.~Bateman, \href{https://arxiv.org/abs/2011.11972}{{Foundations of the Socio-physical Model of Activities (SOMA) for Autonomous Robotic Agents}} (2020).
\newblock \href {http://arxiv.org/abs/2011.11972} {\path{arXiv:2011.11972}}.
\newline\urlprefix\url{https://arxiv.org/abs/2011.11972}

\bibitem{kunze2018soma}
L.~Kunze, H.~Karaoguz, J.~Young, F.~Jovan, J.~Folkesson, P.~Jensfelt, N.~Hawes, {{SOMA}: A Framework for Understanding Change in Everyday Environments using Semantic Object Maps}, Proceedings of the AAAI Fall Symposium on Reasoning and Learning in Real-World Systems for Long-Term Autonomy (LTA) (2018) 47--54.

\bibitem{bredereke2025modularpipeline3dobject}
L.~Bredereke, Y.~Hartmann, T.~Schultz, \href{https://arxiv.org/abs/2503.04322}{{A Modular Pipeline for 3D Object Tracking Using RGB Cameras}} (2025).
\newblock \href {http://arxiv.org/abs/2503.04322} {\path{arXiv:2503.04322}}.
\newline\urlprefix\url{https://arxiv.org/abs/2503.04322}

\bibitem{hartmannkloeckner}
Y.~Hartmann, R.~E. Paul, J.~Klöckner, L.~Deichsel, T.~Schultz, \href{https://doi.org/10.1177/27723577251320237}{{Gait Parameter Estimation from a Single Depth Sensor}}, Journal of Smart Cities and Society 4~(1) (2025) 35--61.
\newblock \href {http://arxiv.org/abs/https://doi.org/10.1177/27723577251320237} {\path{arXiv:https://doi.org/10.1177/27723577251320237}}, \href {https://doi.org/10.1177/27723577251320237} {\path{doi:10.1177/27723577251320237}}.
\newline\urlprefix\url{https://doi.org/10.1177/27723577251320237}

\bibitem{Richter_Putze_Ivucic_Brandt_Schütze_Reisenhofer_Wrede_Schultz_2023}
B.~Richter, F.~Putze, G.~Ivucic, M.~Brandt, C.~Schütze, R.~Reisenhofer, B.~Wrede, T.~Schultz, {EEG Correlates of Distractions and Hesitations in Human–Robot Interaction: A LabLinking Pilot Study}, Multimodal Technologies and Interaction 7~(437) (2023).
\newblock \href {https://doi.org/10.3390/mti7040037} {\path{doi:10.3390/mti7040037}}.

\end{thebibliography}





\end{document}